\documentclass[twocolumn,amsmath,amssymb,nofootinbib,eqsecnum,tighten,floatfix, 
prb]{revtex4}

\usepackage{graphicx}
\usepackage{dcolumn} 
\usepackage{bm, bbm}      
\usepackage{curves}
\usepackage{epic}
\usepackage{wasysym}
\usepackage{pslatex} 
\usepackage{latexsym}
\usepackage[mathcal]{euscript}
\usepackage{epsf,times}
\usepackage{subfigure}

\newcommand{\ket}[1]{\left|#1\right\rangle}

\newtheorem{lem}{Lemma}
\newtheorem{thm}{Theorem}

\begin{document}
\title{Exactly solvable 3D quantum model with finite temperature
topological order}
\author{Isaac H. Kim}
\affiliation{Institute of Quantum Information, California Institute of Technology, Pasadena CA 91125,
USA}

\date{December 3, 2010}

\begin{abstract}
We present a family of exactly solvable spin-$\frac{1}{2}$ quantum hamiltonians on a 3D lattice. The degenerate ground state of the system is characterized by a quantum error correcting code whose number of encoded qubits are equal to the second Betti number of the manifold. These models 1) have solely local interactions, 2) admit a strong-weak duality relation with an Ising model on a dual lattice 3) have topological order in the ground state, some of which survive at finite temperature, 4) behave as classical memory at finite temperature. The associated quantum error correcting codes are all non-CSS stabilizer codes. 
\end{abstract}

\maketitle

\section{Introduction}
One of the motivations for studying quantum error correcting code on lattice is to protect quantum information without active correction. Many models on 2D lattices have been proposed and analyzed ~\cite{Kitaev2003,Bombin2006,Kitaev2006,Fendley2005,Ioselevich2002,Levin2005,Moessner2001} but no-go theorem rules out all finite-range finite-strength hamiltonian system in 2D as a self-correcting quantum memory.\cite{Bravyi2009,Bravyi2008} This does not apply to higher dimensions. For instance, it was shown that 4D toric code is a self-correcting quantum memory.\cite{Dennis2002,Alicki2008} Bombin et al. showed that there is also a 6D model that exhibits similar behavior.\cite{Bombin2009} Whether such thermally protected model exists in 3D remains as an open problem. 3D toric code can store classical information at finite
temperature but it fails to do so for quantum information.\cite{Castelnovo2008} Toplogical color code in 3D, albeit
lacking a rigorous proof, is believed to show a similar behavior: there exists a string-like logical operator which is
thermally unstable.\cite{Bombin2007} 3D model proposed by Nussinov and Ortiz shows similar behavior.\cite{Nussinov2007,Nussinov2008} Another model was proposed by Chamon and analyzed recently by Bravyi et al. This model may be able to protect quantum information, but not in a thermodynamic sense.\cite{Chamon2005,Bravyi2010}

It is worth noting that all the listed 3D models except Chamon's model share a similar property: the quantum error correcting code defining the ground state of the system is a CSS code, meaning that it can be decomposed into two classical codes. When studying the stability of these models, one can show that one of the codes can protect classical information from thermal fluctuation while the other one cannot. This means that there is a manifest difference between how the models treat the bit flip error and the phase flip error. Chamon's model treats $X$, $Y$, and $Z$ error in an identical manner but it lacks stability in thermal sense. Since we expect a singular behavior at the phase boundary between an `ordered state' and `disordered state' for thermally stable quantum memory, absence of finite-temperature phase transition seems troublesome unless there is an argument that can evade this logic. Motivated by these ideas, we present a new spin-$\frac{1}{2}$ model with finite temperature phase transition whose ground state is a non-CSS quantum error correcting code. Our model exhibits a topological order, but only the classical part survives in finite temperature.

The outline of the paper is as follows. We set the stage by introducing the hamiltonian in Section \ref{sec:Model}. In Section  \ref{sec:QuantumCode}, we study the quantum code that defines the ground state of the hamiltonian. We calculate the number of qubits and find the logical operators. In Section \ref{sec:Excitation}, we study the 
low-energy excitation of the hamiltonian that consists of particles and closed strings.  We construct a duality relation with classical Ising model in Section \ref{sec:Duality} to show the finite temperature phase transition.

\section{Model}\label{sec:Model}
  We place qubits on a vertices of a 4-valent 3D lattice. Using the notation $X_i\equiv \sigma_{i}^{x}$, $Y_i \equiv \sigma_{i}^{y}$, $Z_i \equiv \sigma_{i}^{z}$ stabilizer generators are
\begin{align}
   B_{p}^{x} &= \Pi_{i\in p} X_i \\
B_{p}^{y} &= \Pi_{i\in p} Y_i \\
B_{p}^{z} &= \Pi_{i\in p} Z_i,
\end{align}
where $p$ is the plaquette and $\{i \in p\}$ denotes a set of vertices on plaquette $p$. We shall partition a set of plaquettes into $P_x,P_y,P_z$, which corresponds to a set of nontrivial supports for $B_{p}^x, B_{p}^{y},
B_{p}^{z}$. We shall call elements of these sets as $X-,Y-,Z-$plaquettes.

Our model is inspired by the construction of topological color code in 3D.\cite{Bombin2007} For this quantum code, qubits reside on the vertices of the lattice, and the lattice is locally 4-valent. The stabilizer
generators are either a product of $X$s or product of $Z$s, and they correspond to the unit cells of different dimensions; in one example, generators are either in cubic form or plaquette form. Our approach differs in a sense
that we only allow plaquette operators as stabilizer generators.

\begin{figure}[h]
\centering
\subfigure[Vertex Figure]{\label{fig:VertexFigure}
\includegraphics[width=0.2\textwidth]{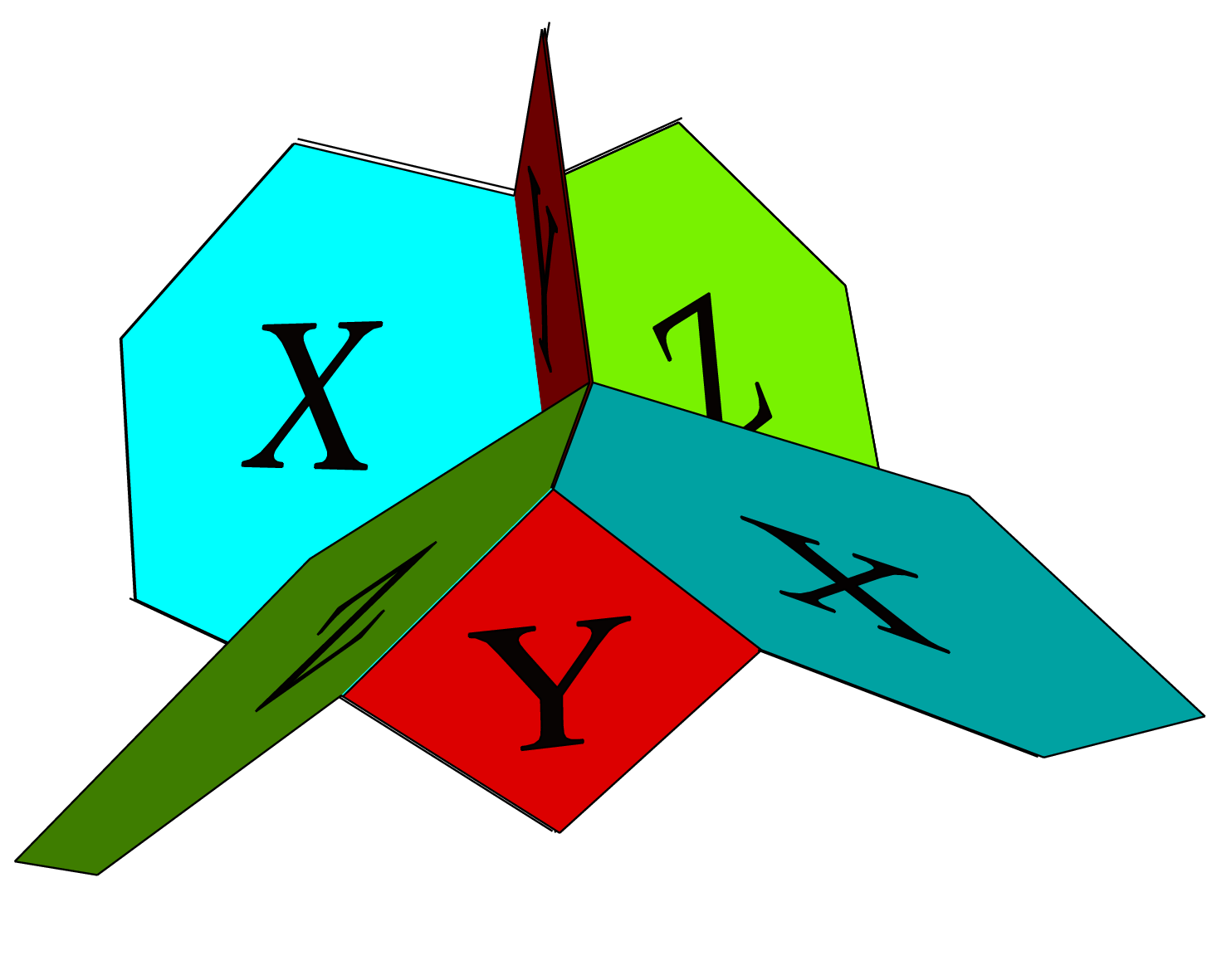}}
\subfigure[Unit Cell]{\label{fig:UnitCell}
\includegraphics[width=0.2\textwidth]{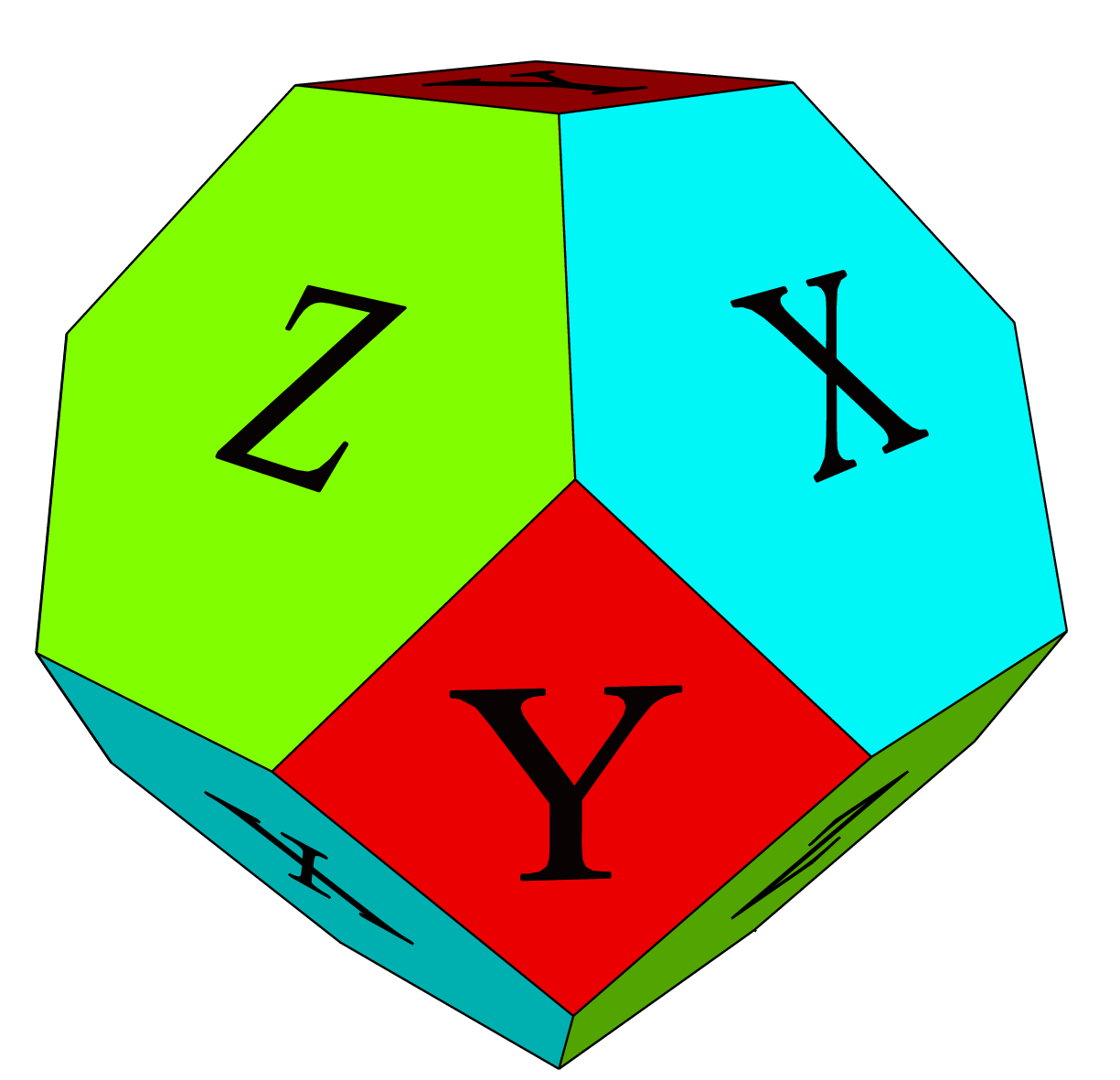}}
    \caption{Vertex figure and unit cell of our model. Qubits reside on the
vertices. One can see that $B_{p_x}^x$ meets with another $B_{p_x}^x$
at one vertex whereas it meets with $B_{p_y}^y$ and $B_{p_z}^z$ at two
vertices.}
\end{figure}

\begin{figure}[h]
\centering
\includegraphics[width=0.3\textwidth]{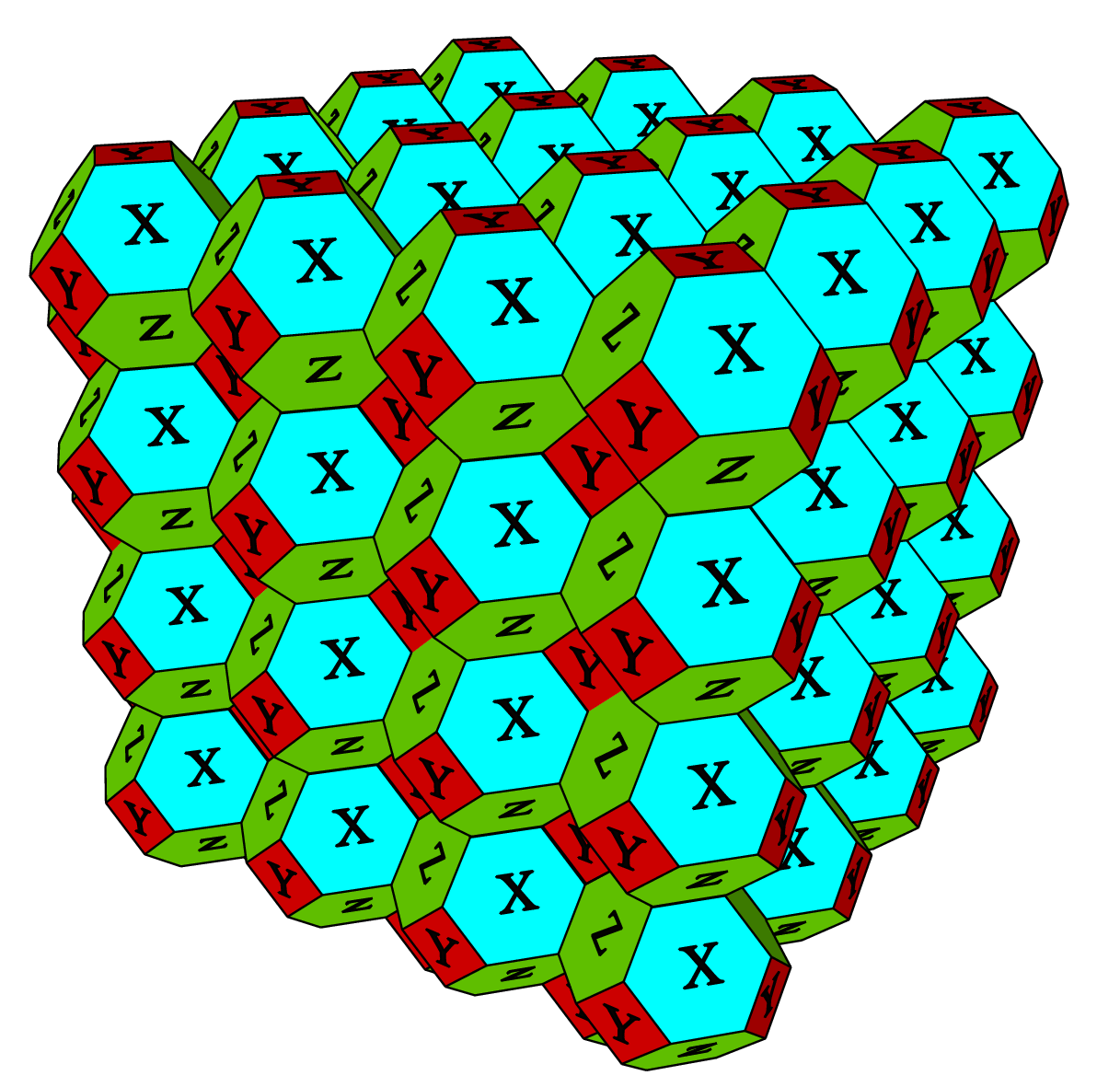}
\caption{Arrangement of stabilizer generators. Translation of unit cells
form a tessellation.\label{fig:OurModel}}
\end{figure}

Local description of our model can be seen in  FIG.\ref{fig:VertexFigure}. At each vertex, there are 6 plaquette operators that have nontrivial support on it. Each plaquette operators meet with a same kind of plaquette operator on
each vertices and meet with $4$ other plaquette operators on $2$ vertices. Thus the assignment in FIG.\ref{fig:VertexFigure} guarantees commutativity between the stabilizer operators.  We must point out that not every lattice structure allows vertex figure like FIG.\ref{fig:VertexFigure}. There are only $4$ translationally invariant convex tessellations that have tetrahedral vertex figure: bitruncated qubic honeycomb,
cantitruncated cubic honeycomb, omnitruncated cubic honeycomb, and cantitruncated alternated cubic honeycomb.\cite{Grunbaum1994} Only the first three admits an arrangement of plaquette operators similar to
FIG.\ref{fig:VertexFigure} at every vertex. In this paper, we mainly study the bitruncated qubic honeycomb model for its simplicity but analogous results shall be discussed in full generality if possible. Unit cell is
shown in FIG.\ref{fig:UnitCell} and tessellation is shown in FIG.\ref{fig:OurModel}. Bitruncated qubic honeycomb is a space-filling  tessellation made up of truncated octahedra. It has 14 faces, 36 edges, and 24 vertices. There are 6 square faces and 8 hexagonal faces. Without  loss of generality, one can set the $6$ square faces to be $Y$ plaquette operator, $4$ of the hexagonal faces to be $X$ plaquette operator and $4$ remaining hexagonal faces to be $Z$ plaquette operators.Hamiltonian is a sum over the  plaquette operators.
\begin{equation}
H= -J (\sum_{p_x \in P_x} B_{p_x}^x +\sum_{p_y \in P_y} B_{p_y}^y +
\sum_{p_z \in P_z} B_{p_z}^z).
\label{eq:Hamiltonian}
\end{equation}

\section{Quantum Code} \label{sec:QuantumCode}
Purpose of this section is to study the quantum code generated by a
set of group generators $\{B_{p_x}^x, B_{p_y}^y, B_{p_z}^z \}$.  The
section is mainly divided into two parts. In Section
\ref{sec:GeometricInterpretation}, we count the number of encoded qubits.
In Section \ref{sec:LogicalOperator}, we completely specify a set of
logical operators for each qubits.

\subsection{Number of Encoded Qubits} \label{sec:GeometricInterpretation}
   Number of encoded qubits can be computed from the size of the
stabilizer group and the number of physical qubits. Since the plaquette
operators are not independent to each
other, we must count the number of independent relations. In such
pursuit, geoemetrical interpretation of our model becomes useful. We
would first like to point out that multiplying all the plaquette
operators on a unit cell reduces to identity. One can see this from
FIG.\ref{fig:UnitCell}. Since any contractible closed surface on the
lattice can be represented as a union of unit cells, one can see that
multiplication of plaquette operators on \emph{any} contractible closed
surface reduces to identity. Therefore we have $C-1$ independent relations
which generate smooth deformation, where $C$ is the number of 3-cells. We
must subtract $1$ becuase multiplying all but one cell results in a
relation for that very cell.

\begin{figure}[h!]
\centering
\subfigure[Top View]{\label{fig:Constraint_TopView}
\includegraphics[width=0.2\textwidth]{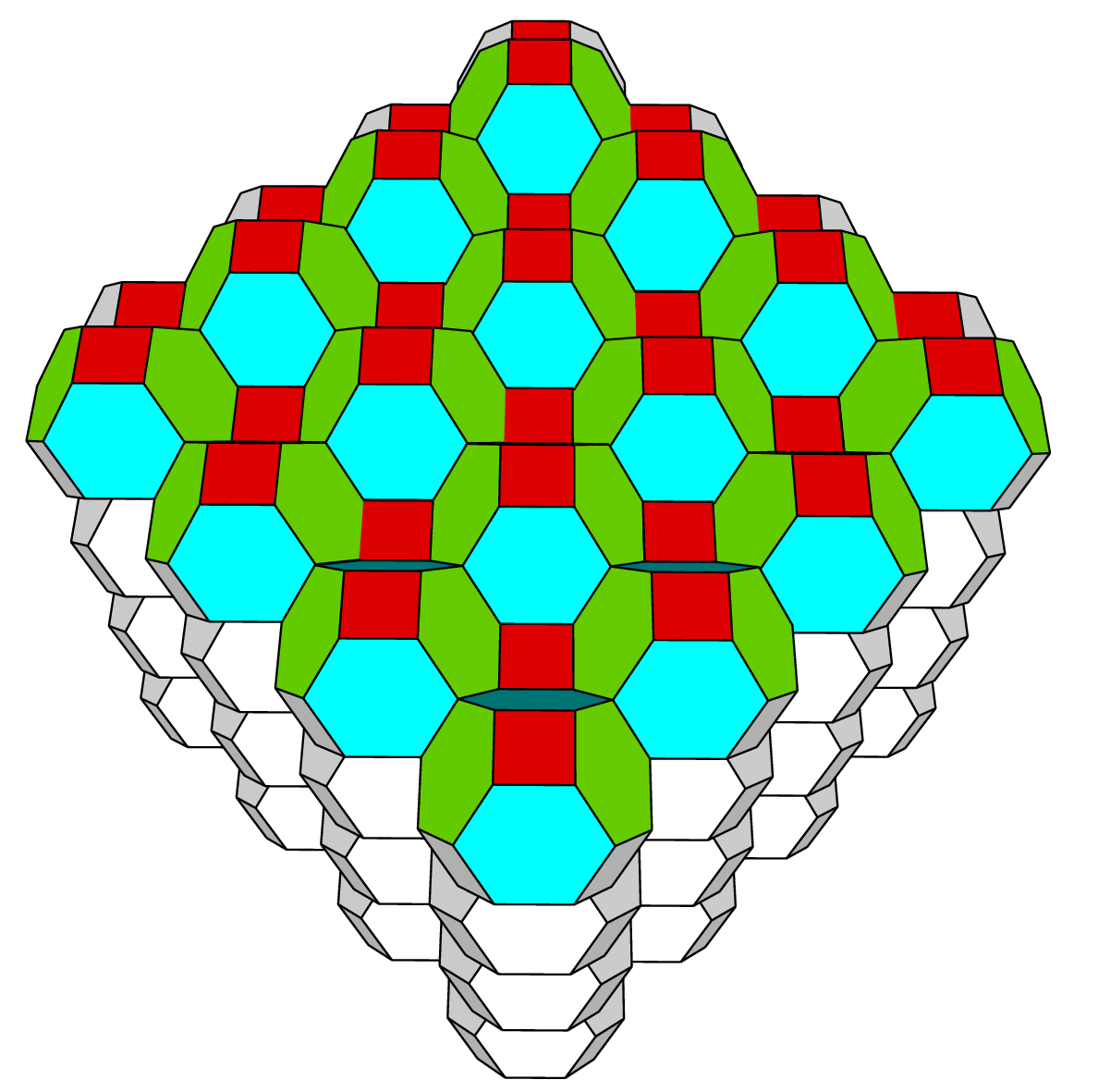}}
\subfigure[Side View]{\label{fig:Constraint_SideView}
\includegraphics[width=0.2\textwidth]{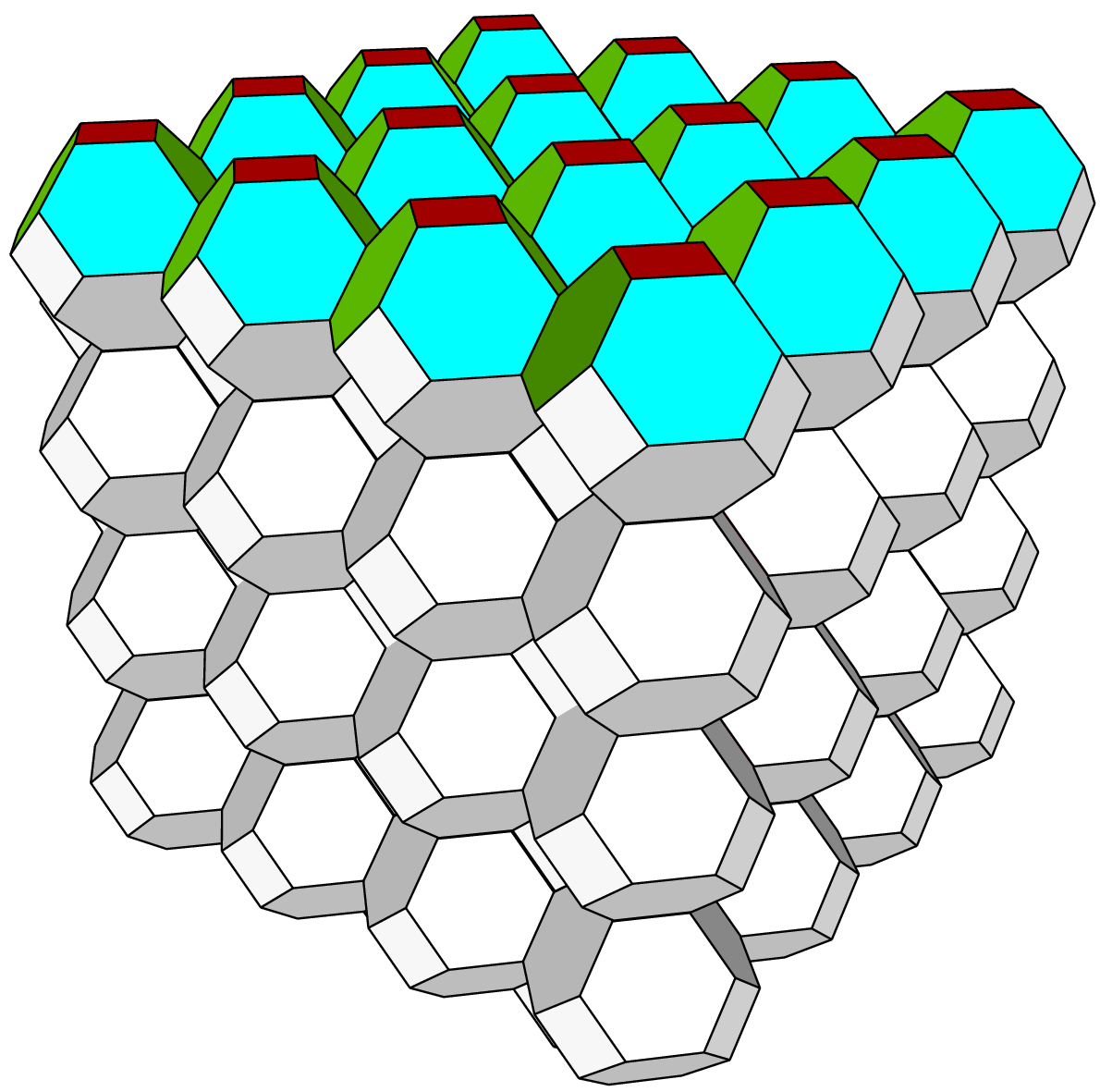}}
\caption{Representation of nontrivial constraints between the stabilizer
operators. One can see that multiplication of all the plaquette operators
on a noncontractible closed surface reduces to identity. At each vertex,
there are either 1) exactly one X, one Y, and one Z or 2) two $X$s and 
two $Z$s.}
\end{figure}

Let us consider a periodic boundary condition on all $3$ directions. There
exists noncontractible surface that reduces to identity as one can see in
FIG.\ref{fig:Constraint_TopView}, FIG.\ref{fig:Constraint_SideView}. Since
there are $3$ topologically distinct noncontractible surfaces, we have $3$
independent relations, resulting in $C+2$ independent relations. Finally,
multiplying all $X$-like operators adds one independent relation. One can
check that multiplication of $Y$s and multiplication of $Z$s are implied
by the previously mentioned relations.

Accounting for these relations, number of encoded qubits is $V-F+C+3=3$.
This reasoning can be generalized to any orientable 3-manifold.

\begin{lem}
   For stabilizer group $\{B_{p_x}^x, B_{p_y}^y, B_{p_z}^z \}$, $k=b_2$.
\end{lem}
\textbf{Proof :} We use the definition of Euler Characteristic.
\begin{equation}
   \chi = V-E+F-C = 0
\end{equation}
$\chi$ is trivially $0$ due to Poincar\'e Duality. In the dual lattice,
$V$ is the number of tetrahedral cells. $E$ is number of
faces, and hence $E=2V$. Therefore we have
\begin{equation}
   V-(F-C)=0.
\end{equation}
Hence
\begin{align}
   k&=V-(F-(C-1+1+b_2))\\
&=b_2,
\end{align}
where $b_2$ is the second Betti number of the manifold. One can also use
this intuition to prove that the group generated by the plaquette
operators does not contain $-I$.

\begin{lem}
  $\langle B_{p_x}^x, B_{p_y}^y, B_{p_z}^z \rangle$ does not contain $-I$.
\end{lem}
\textbf{Proof:} Any constraint between the plaquette operators can be
represented as a product of closed $3$-manifold. For each unit cell, we
have $24$ vertices at which $X,Y,$ and $Z$ meets. Since all the generators
commute with each other, we can arrange the product to be the following
canonical form.
\begin{equation}
  \Pi_{p_x} B_{p_x}^x \Pi_{p_y} B_{p_y}^y \Pi_{p_z} B_{p_z}^z.
\end{equation}
Since $XYZ=i$, the product of plaquette operators on a unit cell is $1$.
Similarly, product of plaquette operators on a noncontractible surface
described in FIG.\ref{fig:Constraint_TopView},
FIG.\ref{fig:Constraint_SideView}, we have $4n$ vertices where $X,Y,$ and
$Z$ meets. Hence we arrive at the same conclusion. Since any product of
plaquette operators that results in a trivial operator can be constructed
by these constraints, the group does not contain $-I$.

\subsection{Logical Operators} \label{sec:LogicalOperator}
\begin{figure}[h!]
\centering
\includegraphics[width=0.4\textwidth]{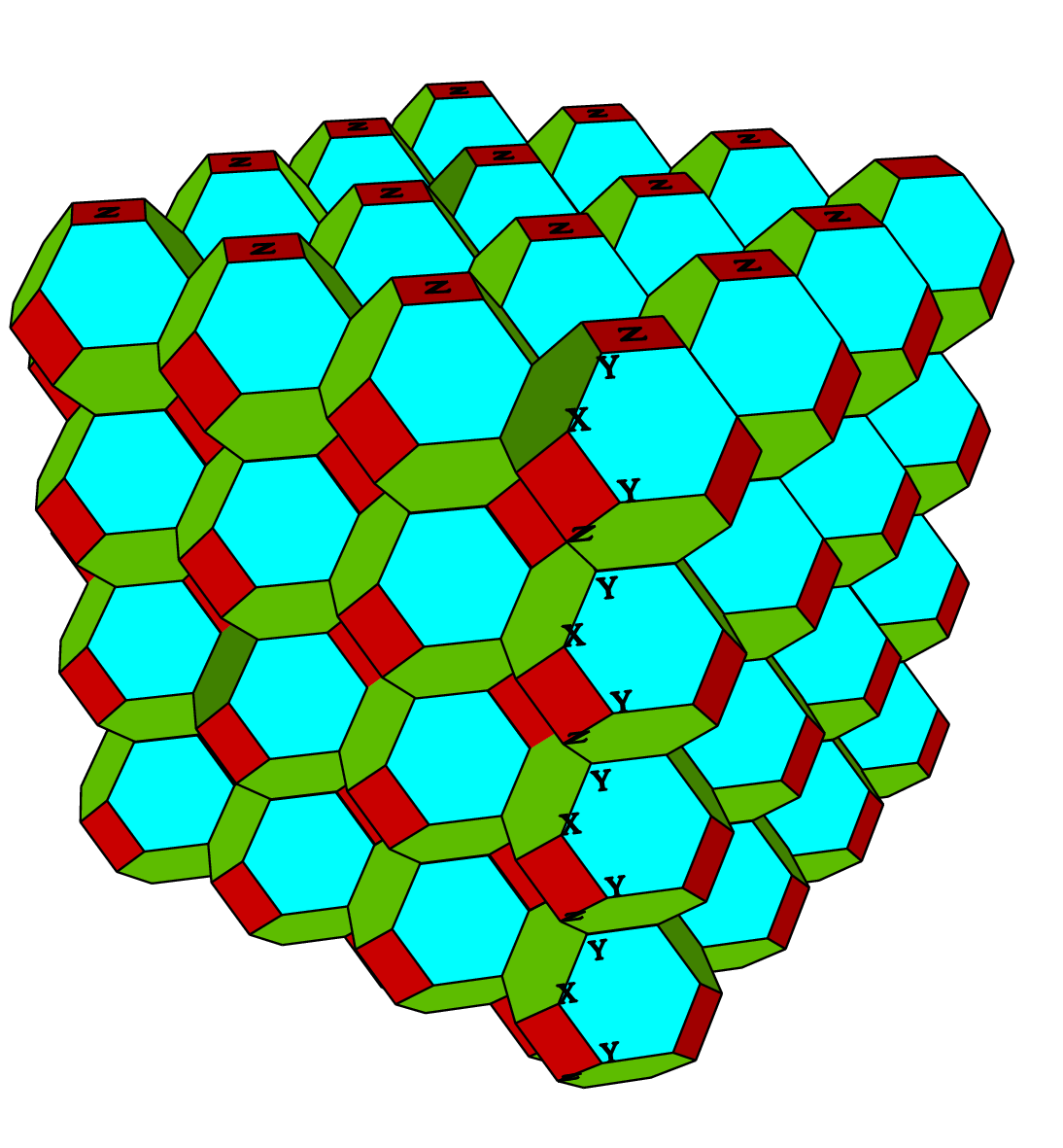}
\caption{There is one surface operator and one string operator for each
qubits. Surface operator corresponds to the product of $ZZZZ$ on 
$Y$-plaquettes. String operator is the line perpendicular to this 
surface, showing a sequence $YZYXYZYX\cdots$. \label{fig:LogicalOperators}}
\end{figure}

There are two logical operators that are reminiscent to the surface and
string operator of 3D toric code. These are drawn in
FIG.\ref{fig:LogicalOperators}. One can see the surface operator on the
top of the lattice system which is a product of $B_{p_y}^z$s on one layer
of $Y$-plaquettes. The complementary logical operator to this is the
string operator that has a sequence of $YZYXYZYXYZYX\cdots$ along the line
perpendicular to the surface operator. This string winds around the torus
and completes a noncontractible loop. These two operators anticommute with
each other and both of them commute with the
stabilizer generators.
We can similarly define two sets of complementary operators in other
directions. One can easily check the expected commutation and
anticommutation relations.

\section{Low energy excitation}\label{sec:Excitation} Quasiparticles 
excitations in 2D typically arise as anyons. For instance, in Kitaev's toric code, two quasiparticles are created in pair, and when fused together, they vanish.\cite{Kitaev2003} There are two kind
of particles analogous to electric and magnetic charge, and when one
particle winds around another one, the system attains a nontrivial global
phase. In 3D, trajectory of winding around another particle can be
deformed into a trivial contour. Hence one needs higher dimensional object
to attain a similar topological action. In 3D there are closed string-like
excitations and particle-like excitations.\cite{Bombin2006,Castelnovo2008}
When the particle winds around the string so that the trajectory and the
string together forms a knot, the system attains a nontrivial global
phase.

Our model presents a similar picture. Particle-like excitations are
created in pair. If we truncate a string-like logical operator,
excitations form at the end points. When the particle-antiparticle pair is
created, they can diffuse without any extra energy cost. Closed
string-like excitations can be similarly thought as a truncated
surface-like logical operator. Near the boundary of the surface, there are
excitations and hence the energy cost grows linearly with the size of the
surface. When a particle penetrates the closed string, we find that
\begin{figure}[hb]
\includegraphics[width=0.4\textwidth]{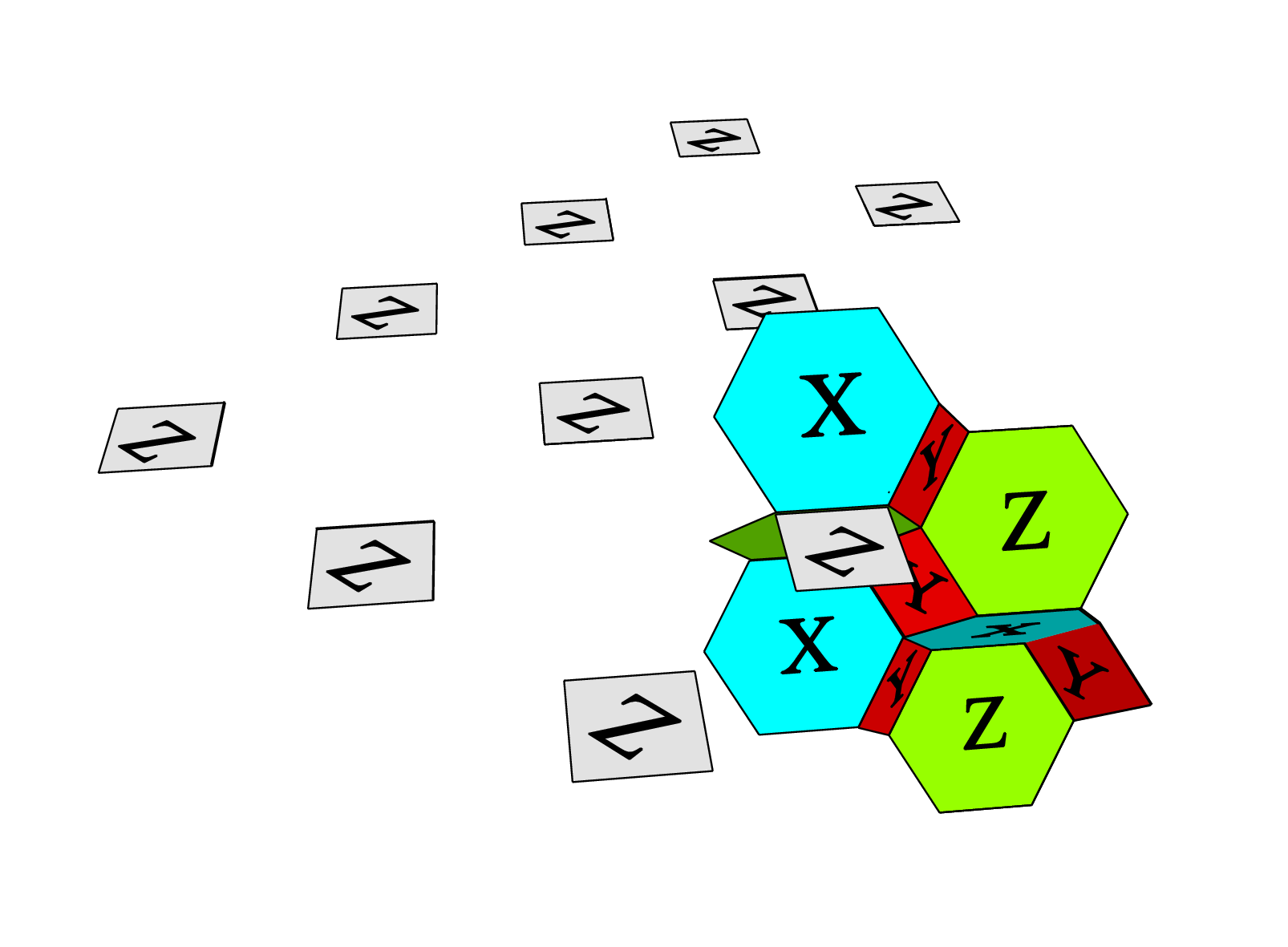}
\caption{Representation of particle penetrating through a string-like
excitation. Truncated surface operator is a product of $Z$-plaquettes in
white. Trajectory of the particle is a nontrivial support of the colored
plaquette operators, which coincides with the $Z$-surface.
\label{fig:ParticleMovement}}
\end{figure}

\begin{align}
\ket{\psi_{Initial}} &= SP \ket{\Phi} \\
\ket{\psi_{Final}} &=  USP \ket{\Phi} = -\ket{\psi_{Initial}},
\end{align}
where $S$ is a closed-string excitation, $P$ is a particle excitation,
and $U$
is a trajectory of the particle. Thus system gains $e^{i\pi}$ phase
factor. This is illustrated in FIG.\ref{fig:ParticleMovement}. One can see
that as a particle penetrates through the surface operator and returns to
the original position, it coincides with the surface operator at one
vertex, thus giving the anticommutation relation.

Low energy excitation in terms of elementary objects provides us an
intuitive picture for the thermal stability. Particles can be created out
of vacuum in pair and propagate freely. They can diffuse and wind around
the torus to induce logical error. Closed strings, on the other hand, need
energy that is proportional to its perimeter. Given a closed string-like
excitation as in FIG.\ref{fig:ParticleMovement}, the stabilizer generators
anticommuting with the surface operator only reside near the boundary of
the surface. $Z$-plaquettes trivially commute with the surface
operator. $X$-plaquettes commute with the surface operator since they meet
at two vertices. However, there are $Y$-plaquettes meeting at exactly one
vertex at the boundary. Hence we expect our system to be a stable
classical memory.

\section{Duality}\label{sec:Duality}
Typical strong-weak duality relation relates a strong coupling limit of
one model to a weak coupling limit of another model: we use a slightly
different strategy here. We first show that our model can be mapped into
an Ising gauge theory, from which we can use the Wegner-type duality
relation with Ising model. Mapping from our model to Ising gauge theory is
not exact for finite sized lattice, but this difference vanishes in the
thermodynamic limit. Starting from the partition function of our model,
\begin{align}
   Z&= tr(\exp(-\beta H))\\
&= tr(\Pi_{S_i \in S}(\cosh \beta J + S_i \sinh \beta J)),
\end{align}
where $S_i \in \{B_{p_x}^x, B_{p_y}^y, B_{p_z}^z \}$,

\begin{align}
Z&= (\cosh \beta J)^n tr(\Pi_i (1+\alpha S_i)) \\
&= (\cosh \beta J)^n tr(\sum_{\{k_i \}=0}^{1} \Pi_i \alpha^{k_i} S_i^{k_i}).
\end{align}

Since the Pauli operators are traceless, the nonvanishing terms correspond
to the nontrivial constraints presented in Section
\ref{sec:GeometricInterpretation}. Note that there were two kind of
constraints: constraints coming from the closed 2-manifold and constraints
coming from space-filling products of $X$, $Y$s, or $Z$s. Using this, we
can write down the partition function in the following form.
\begin{widetext}
\begin{align}
  Z&= (2\cosh \beta J)^n (\sum_{c} \alpha^{A_c}+(1+\alpha^{n_x})(1+
\alpha^{n_y})(1 + \alpha^{n_z})-1  + C.T.)
\end{align}
\end{widetext}

$\sum_c$ is a sum over a configuration of closed $2$-manifolds. $A_c$ is
the number of plaquettes for each configurations. $C.T.$ corresponds to
the cross terms between closed 2-manifolds and space-filling product of
$X$s, $Y$s, or $Z$s. $n_{x,y,z}$ corresponds to the number of
$X,Y,Z-$plaquette operators. The main idea is that the partition function
is dominated by the first term in the thermodynamic limit. We show this in
Appendix \ref{Appendix:cross_term_bound}.

\begin{lem}
  $Z - C.T. - (\alpha^{n_x} + \alpha^{n_y} + \alpha^{n_z}) = Z_{IG}(\beta
J)$, where $Z_{IG}$ is a partition function of Ising gauge theory on the
same lattice with temperature $\beta$ and coupling constant $J$.
\end{lem}
\textbf{Proof : } Consider a mapping $B_{p_x}^x \to ZZZZZZ$, $B_{p_y}^y
\to ZZZZ$, $B_{p_z}^z \to ZZZZZZ$, where $Z\cdots Z$ are products of $Z$
on the edges of each plaquettes. The resulting model is an Ising gauge
theory on a bitrucated cubic honeycomb. Partition function is
\begin{align}
  Z_{IG} &= tr(\exp(-\beta H))\\
&= (\cosh \beta J)^n tr(1+ \tanh \beta J S_i),
\end{align}
where $S_i$s are either $ZZZZZZ$ or $ZZZZ$ depending on the plaquette.
Since Pauli operators are traceless, only a product of plaquette operators
that are union of closed surface survives. Therefore, we conclude
\begin{equation}
  Z_{IG}(\beta J) =Z - C.T. - (\alpha^{n_x} + \alpha^{n_y} + \alpha^{n_z}).
\end{equation}

Using the duality relation between Ising gauge theory and Ising model, we
can map our model into an Ising model. We show the duality relation in
Appendix \ref{Appendix:Duality}.
\begin{thm}
   Our model with coupling constant $\beta J$ is dual to the classical
Ising model on a dual lattice with a dual coupling constant
$\tilde{\beta J} = -\frac{1}{2}
\ln \tanh \beta J$.
\end{thm}
Since the Ising model undergoes a finite temperature phase transition, so
does our model. This is analogous to the behavior of 3D toric code under
temperature change. As in our model, one can show that 3D toric code has
critical temperature by using the duality relation with Ising model. Below
the critical temperature, there is a symmetry breaking with respect to a
surface-like logical operator. Symmetry associated to the string-like
logical operator is broken only at the ground state.

One glaring difference though, is that 3D toric code can be decomposed
into two classical hamiltonians without spoiling the phase transition: the
hamiltonian responsible for correcting the bit flip error is identical to Ising
gauge theory, which has finite temperature phase transition. On the other
hand, the hamiltonian responsible for correcting the phase flip error does
not have a phase transition. Hence one can intuitively understand that 3D
toric code can only correct bit flip errors but not phase flip errors
under thermal equilibrium. Our model does not allow such decomposition.
Once we get rid of any of $B_{p_x}^x, B_{p_y}^y,$ or $B_{p_z}^z$, the
partition function does not exhibit a phase transition any more. This shows that non-CSS code with finite temperature phase transition in 3D does not necessarily provide a self-correcting quantum memory.

\section{Conclusion}
In this paper, we studied an exactly solvable 3D spin model and studied
its topological order. The ground state of the system defines a non-CSS
quantum error correcting code. At finite temperature, this system is
expected to behave as a stable classical memory, but not as a stable
quantum memory. This is mainly due to the fact that there exists a
string-like logical operator. In light of studying the possibility of
self-correcting quantum memory, this reconfirms the general properties
that have been found in  3D stabilizer codes so far: for each encoded
qubit, there exists one surface-like logical operator and one string-like
logical operator. It seems that we cannot avoid such outcome unless the
shape of the logical operator changes as the system size changes, as in
Chamon's model.\cite{Chamon2005,Bravyi2010} This in fact was recently argued to be the general feature of stabilizer codes whose number of encoded qubits remain invariant under system size change. \cite{Yoshida2010}

It is worth noting that the thermal stability analysis of our model is not
rigorous at this stage, even though the energy barrier increasing as the
perimeter of the surface is a compelling evidence that this must be true.
It would be desirable to make a rigorous estimate of thermal relaxation
rate using the method introduced by Chesi et al.\cite{Chesi2009} We expect
the string-like logical operator to be thermally fragile and the
surface-like logical operator to be stable. As in 3D toric
code,\cite{Castelnovo2008} we also expect the topological entropy of our
model to show a singular behavior near the critical point. These singular
behavior arise due to the existence of finite temperature phase
transition, which we can show rigorously by the strong-weak duality
relation between our quantum model to a classical Ising model on the dual lattice.

\begin{acknowledgements}
Author would like to thank Jeongwan Haah for his help in finding the logical
operator of the system, and John Preskill for many insightful 
discussions. This research was supported in part by NSF under Grant No. PHY-0803371 and ARO under Grant No. W911NF-09-1-0442.
\end{acknowledgements}

\appendix
\section{Bound for the cross terms.}\label{Appendix:cross_term_bound}

Corss term can be written as
  \begin{align}
   C.T.= \sum_c \alpha^{A_c} \sum_{i\in\{x,y,z \}}\alpha^{n_i - 2n_i^c}  ,
  \end{align}
where $n_x,n_y,n_z$ are total number of $X,Y,Z-$plaquettes and $n_x^c,
n_y^c,n_z^c$ are number of $X,Y,Z-$plaquettes for configuration $c$.

\begin{lem}
There exists $0<\epsilon_{1,2} <1$ such that
  \begin{equation}
   A_c  +n_i-2n_i^c \geq \epsilon_1 A_c + \epsilon_2 n_i
  \end{equation}
for $\forall c, i$.
\end{lem}
\textbf{Proof : }
Consider $i=x$. Left hand side of the inequality is
\begin{align}
n_y^c+n_z^c-n_x^c + n_x &\geq n_y^c + n_z^c - (1-\epsilon)n_x^c  +
(1-\epsilon)n_x \\
&\geq (\frac{\epsilon}{2})A_c  + (1-\epsilon)n_x
\end{align}
On the second line, we used the fact that the minimum is achieved in the case where $n_y^c=0$, implying $n_z^c=n_x^c=\frac{1}{2}A_c$.
Same logic can be applied to $i=z$. For $i=y$,
\begin{align}
n_x^c+n_z^c-n_x^c + n_y &\geq n_x^c + n_z^c - (1-\epsilon)n_y^c  +
(1-\epsilon)n_y \\
&\geq (\frac{2}{5} - \frac{3}{5}(1-\epsilon))A_c  + (1-\epsilon)n_y.
\end{align}
Similarly, here we used the fact that the minimum is achieved in the case where one of $n_x^c$ or $n_z^c$ is $0$. Then we have a $2:3$ ratio between the $X-(Z-)$plaquettes and $Y-$plaquettes. 
Therefore, for $\epsilon > \frac{1}{3}$, we have such $(\epsilon_1,
\epsilon_2)$.

\begin{lem}
\begin{equation}
\lim_{vol \to \infty} \frac{Z(\beta J)}{Z_{IG}(\beta J)} \to 1.
\end{equation},
where $Z_{IG}(\beta J)$ is a partition function for Ising gauge theory
with temperature $\beta$ and coupling constant $J$. $vol$ is the volume of
the lattice.
\end{lem}
\textbf{Proof :}

We use
\begin{align}
  \sum_c \alpha^{\epsilon_1 A_c}&= \frac{(2\cosh \beta J')^n}{(2\cosh \beta
J')^n} \sum_c \alpha'^{A_c} \\
&=(\frac{1}{2\cosh \beta J'})^n Z_{IG}(\beta J'),
\end{align}
where
\begin{equation}
\tanh \beta J' = (\tanh \beta J)^{\epsilon_1}.
\end{equation}

Thus the cross terms can be bound by
\begin{equation}
Z_{IG}(\beta J') (\frac{\cosh \beta J}{\cosh \beta J'})^n\alpha^{\delta_i
\epsilon_2 n},.
\end{equation}
where $\delta_i = \frac{n_i}{n}$,where $n$ is the total number of plaquettes. This becomes

\begin{equation}
  Z_{IG}(\beta J')
((\frac{1-t^2}{1-t^{\frac{2}{\epsilon_1}}})^{\frac{1}{2}}
t^{\frac{\epsilon_2}{\delta \epsilon_1}})^n,
\end{equation}
where $t=\tanh \beta J'$. One can show that
$(\frac{1-t^2}{1-t^{\frac{2}{\epsilon_1}}})^{\frac{1}{2}}
t^{\frac{\epsilon_2}{\epsilon_1 \delta}} < 1$ for $\beta J>0$. Since the
renormalized coupling constant $J'$ is larger than $J$, we can see that
these correction terms become negligible in thermodyamic limit. Therefore,
\begin{equation}
  |\lim_{vol \to \infty} \frac{Z(\beta J) - Z_{IG}(\beta J)}{Z_{IG}(\beta
J)}| \leq
|\frac{Z_{IG}(\beta J') }{Z_{IG}(\beta J)} \lambda^n  + O(\alpha^n)|,
\end{equation}
where $J' > J$ and $0<\lambda<1$. In $n\to \infty$ limit, we get the
desired result.

\section{Duality between Ising gauge theory and Ising
model}\label{Appendix:Duality}
\begin{lem}
Ising gauge theory on bitruncated cubic honeycomb is dual to Ising model
on the dual lattice.
\end{lem}
\text{Proof: }
  \begin{align}
   Z&=(\cosh \beta J)^n tr(\Pi_i (1+\tanh \beta J S_i)) \\
&= (\cosh \beta J)^n tr(\sum_{\{k_i \}=0}^{1} \Pi_i \alpha^{k_i} 
S_i^{k_i} ) \\
&= (2\cosh \beta J)^n \sum_{\{k_i \}=0}^{1} \Pi_i\alpha^{k_i}
\Pi_{e}\delta_2(\sum_{j} k_{j;e}),
  \end{align}
where $\Pi_e$ is a product over all the edges and $\sum_j k_{j;e}$ is a
sum over $k_j$s that have nontrivial support on edge $e$. There are three
such $k_j$s. One can use $k_{j;e}=\frac{1}{2}(1-ZZ)$, where $ZZ$ is a
product of $Z$s on qubits that reside on the vertices of the dual lattice.
For $8$ spin configurations $(Z_1,Z_2,Z_3)=$ $(-1,-1,-1),$ $(1,1,1),$
$(1,-1,-1),$ $(-1,1,-1),$ $(-1,-1,1),$ $(1,1,-1),$ $(-1,1,1),$ $(1,-1,1),$
one can see that all of these configurations satisfy the delta function.
Furthermore, we have 2 combinations for $(k_1, k_2, k_3) = (0,0,0)$, 2
combinations for $(0,1,1)$, $(1,0,1)$, and $(1,1,0)$. Plugging this in, 
we get
\begin{align}
Z&= (\cosh \beta J)^n \sum_{\{Z_i\}=0}^1 \Pi_i 
\alpha^{1-\frac{1}{2}Z_{i+\hat{n}_i}Z_{i - \hat{n}_i}},
\end{align}
where $Z_{i\pm \hat{n}_i}$ is the $Z$ operator on the dual sites of 
plaquette $i$. $\hat{n}_i$ is the unit normal vector to the plaquette.
Therefore, up to a constant, partition function is identical to the 
partition of Ising model with $\tilde{\beta J} = -\frac{1}{2}\ln \tanh \beta J$.

\bibliographystyle{apsrev}
\bibliography{bib}

\end{document}